\documentclass[%
aip,
apl,
 amsmath,amssymb,
 reprint,%
]{revtex4-1}

\usepackage{lineno}
\usepackage[colorlinks=true,linkcolor=blue]{hyperref}
\usepackage[squaren]{SIunits}
\usepackage{color}
\usepackage{amsmath}
\usepackage{setspace}
\usepackage{multirow}
\usepackage{threeparttable}
\usepackage{booktabs}
\usepackage{graphicx}
\usepackage{dcolumn}
\usepackage{bm}
\usepackage{chemformula} 

\usepackage[utf8]{inputenc}
\usepackage[T1]{fontenc}
\usepackage{mathptmx}
\usepackage{etoolbox}

\definecolor{americanrose}{rgb}{1.0, 0.01, 0.24}
\definecolor{coralpink}{rgb}{0.97, 0.51, 0.47}
\definecolor{ao(english)}{rgb}{0.0, 0.5, 0.0}
\definecolor{darkpastelgreen}{rgb}{0.01, 0.75, 0.24}
\definecolor{cyan(process)}{rgb}{0.0, 0.72, 0.92}

\usepackage{epstopdf}

\definecolor{cream}{RGB}{222,217,201}

\makeatletter
\def\@email#1#2{%
 \endgroup
 \patchcmd{\titleblock@produce}
  {\frontmatter@RRAPformat}
  {\frontmatter@RRAPformat{\produce@RRAP{*#1\href{mailto:#2}{#2}}}\frontmatter@RRAPformat}
  {}{}
}%
\makeatother
\begin{document}

\preprint{AIP/123-QED}

\title[]{Effects of Structural Variations to X-ray Absorption Spectra of g-C$_3$N$_4$: Insights from DFT and TDDFT Simulations}
\author{Jun-Rong Zhang}
 \affiliation{MIIT Key Laboratory of Semiconductor Microstructure and Quantum Sensing, Department of Applied Physics, School of Physics, Nanjing University of Science and Technology, 210094 Nanjing, China.}
\author{Sheng-Yu Wang}%
 \email{wangshengyu@njust.edu.cn}
 \affiliation{MIIT Key Laboratory of Semiconductor Microstructure and Quantum Sensing, Department of Applied Physics, School of Physics, Nanjing University of Science and Technology, 210094 Nanjing, China.
}%
\author{Minrui Wei}
\affiliation{MIIT Key Laboratory of Semiconductor Microstructure and Quantum Sensing, Department of Applied Physics, School of Physics, Nanjing University of Science and Technology, 210094 Nanjing, China.
}%
\author{Qiang Fu}
\affiliation{Hefei National Laboratory for Physical Sciences at the Microscale, University of Science and Technology of China, 230026 Hefei, China.
}%
\author{Weijie Hua}
\affiliation{MIIT Key Laboratory of Semiconductor Microstructure and Quantum Sensing, Department of Applied Physics, School of Physics, Nanjing University of Science and Technology, 210094 Nanjing, China.
}%

\date{\today}
             
\begin{abstract}
X-ray absorption spectroscopy (XAS) is widely employed for structure characterization of graphitic carbon nitride (g-C$_3$N$_4$) and its composites. Nevertheless, even for pure g-C$_3$N$_4$, discrepancies in energy and profile exist across different experiments, which can be attributed to variations in structures arising from diverse synthesis conditions and calibration procedures. Here, we conducted a theoretical investigation on XAS of three representative g-C$_3$N$_4$ structures (planar, corrugated, and micro-corrugated) optimized with different strategies, to understand the structure-spectroscopy relation. Different methods were compared, including density functional theory (DFT) with the full (FCH) or equivalent (ECH) core-hole approximation, as well as the time-dependent DFT (TDDFT).  FCH was responsible for getting accurate absolute absorption energy; while ECH and TDDFT aided in interpreting the spectra, through ECH-state canonical molecular orbitals (ECH-CMOs) and natural transition orbitals (NTOs), respectively. With each method, the spectra at the three structures show evident differences, which can be correlated to different individual experiments or in between. Our calculations explained the structural reason behind the spectral discrepancies among different experiments. Moreover, profiles predicted by these methods also displayed consistency, so their differences can be used as a reliable indicator of their accuracy. Both ECH-CMOs and NTO particle orbitals led to similar graphics, validating their applicability in interpreting the transitions. This work provides a comprehensive analysis of the structure-XAS relation for g-C$_3$N$_4$, provides concrete explanations for the spectral differences reported in various experiments, and offers insight for future structure dynamical and transient X-ray spectral analyses.
\end{abstract}

\maketitle


X-ray absorption spectroscopy (XAS) is a powerful tool for characterizing material structures. Grounded on core excitations, this technique is sensitive to local bonding structures as well as structural changes.\cite{stohr1992} Understanding the structure-XAS relation is crucial to uncovering the underlying physical and chemical insights.

Graphitic carbon nitride (g-C$_3$N$_4$) has received much research attention in recent years due to its potential usages in photocatalyst, adsorbent, composite membrane, and disinfectant.\cite{zhu_recent_2023, *li_review_2022, *wang_review_2022, *wang_membrane_2022, *xing_review_2021} XAS spectra, both at the C and N K-edges, are widely employed in structure characterizations of g-C$_3$N$_4$,\cite{mehtab_XAS_2022, zhang_precisely_2022, meng_superficial_2019, zheng_high_2016, kamal_hussien_metal-free_2022} its composites (e.g., SnS/g-C$_3$N$_4$,\cite{omr_design_2023} B/g-C$_3$N$_4$,\cite{kamal_hussien_metal-free_2022} and Ni/g-C$_3$N$_4$\cite{wang_attenuating_2022}), and derivatives.\cite{zhang_synergistic_2022, *zhao_mechanistic_2021, *che_iodideinduced_2021} However, even for pure g-C$_3$N$_4$, different experiments\cite{kamal_hussien_metal-free_2022, zheng2014hydrogen, Liu_exp_2015, Che_exp_JACS2017, Lee_exp2014} reported spectra with evident differences, attributed to different synthesis (causing structural differences) and calibration (causing energy shifts) procedures used. Especially, synthesis processes involve different precursors, temperatures, crystallite sizes (partially or fully polymerized), and introduced impurities, leading to spectral variations.\cite{sharma_synthetize_2018, *du_growth_nodate}

\begin{figure}
\centering
\includegraphics[width=8.5cm]{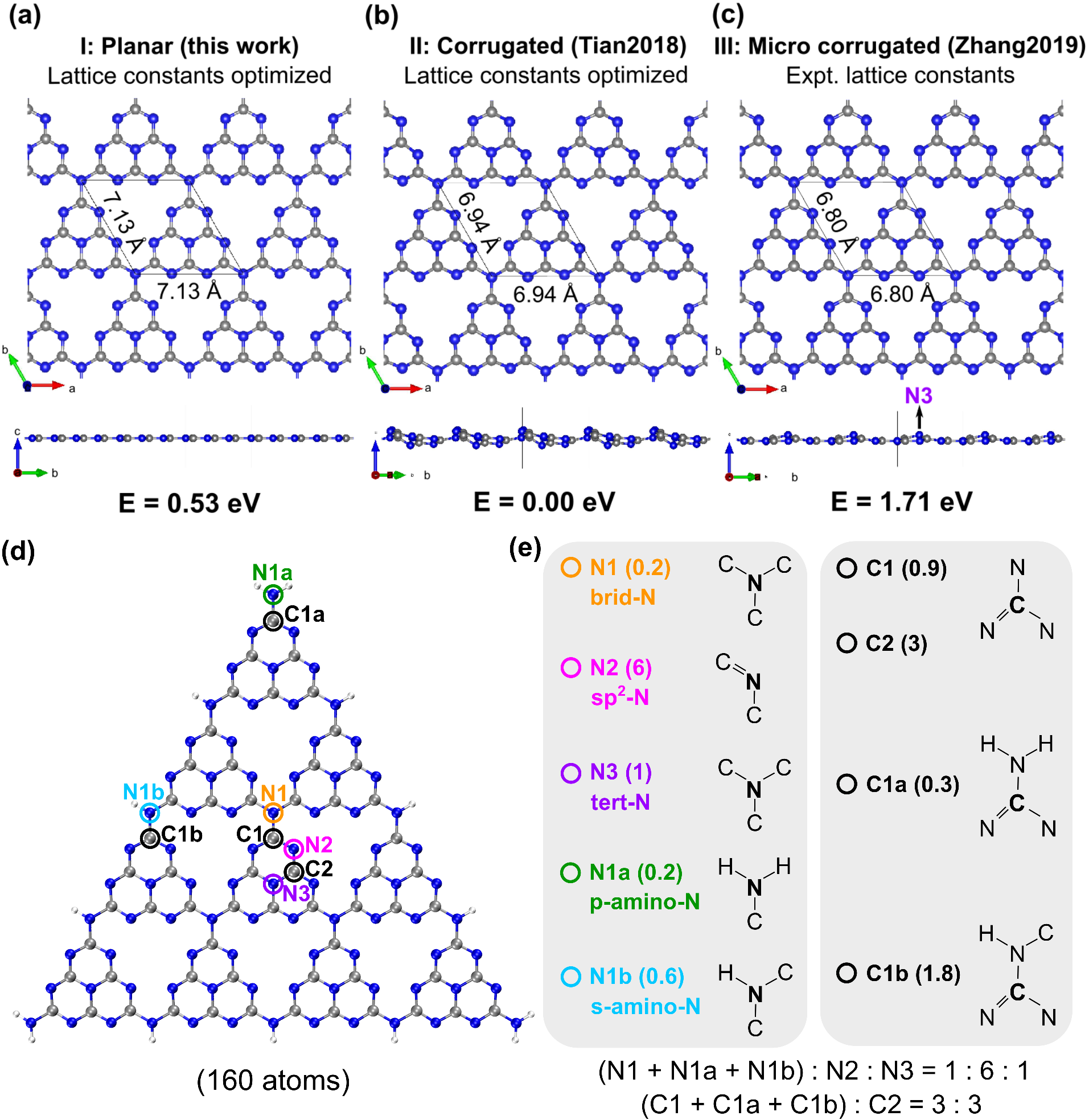}
\caption{(a-c) Three g-C$_3$N$_4$ structures under study: (a)  planar (I), (b)  corrugated\cite{tian_NatCommun_2018} (II), and (c)  micro-corrugated\cite{zhang_accurate_2019} (III, with lattice constant fixed at experimental values\cite{wang_6.81_2009}) conformations. \textit{Top}, top view.  Lattice constants are labeled. \textit{Bottom}, side view (see Fig. S1 for an enlarged view). Simulated relative energies are indicated. (d) Cluster model used for XAS simulations of I-III. Circles indicate different types of N/C atoms used for core excitations. (e) Local bonding structure of each type. Atomic weights used to generate the total XAS are given in parentheses.\cite{zhang_accurate_2019}} 
\label{fig:periodicStructure} 
\end{figure}

There is an open issue of whether the structure of g-C$_3$N$_4$ is \emph{planar}\cite{khamdang_computational_2022, *zhu_tuning_2021, *yang_dft_2021, *liang_7.13_2021, *wang_7.13_2020, *tong_first-principle_2018, *sun_first_2018, *zuluaga_plane_2015} or \emph{corrugated}\cite{negro_combined_2023, liu_buckled_2020, wirth_wave-like_2014, deifallah_electronic_2008, tian_NatCommun_2018, gracia_corrugated_2009} [Fig. \ref{fig:periodicStructure}(a-b); denoted as configurations I and II, respectively]. This issue stems from a discrepancy between the lattice parameters obtained from X-ray diffraction (XRD) patterns and the size of a heptazine unit (explained by small tilt angularity in the structure).\cite{wang_6.81_2009} Theoretical calculations supporting both corrugated\cite{deifallah_electronic_2008, wirth_wave-like_2014, liu_buckled_2020, negro_combined_2023} and planar\cite{khamdang_computational_2022, *zhu_tuning_2021, *yang_dft_2021, *liang_7.13_2021, *wang_7.13_2020, *tong_first-principle_2018, *sun_first_2018, *zuluaga_plane_2015} structures. Particularly, corrugated structures were predicted by density functional theory (DFT) calculations with different functionals: Wirth et al.\cite{wirth_wave-like_2014} obtained a ``wave-like'' sheet with lattice constants of $a$=$b$=6.79 {\AA} by PBE\cite{PhysRevLett.77.3865}; Liu et al.\cite{liu_buckled_2020} optimized a structure with the buckling height of 1.4 {\AA} and $a$=$b$=6.94 {\AA} by HSE03\cite{heyd_hybrid_2003}; Negro et al.\cite{negro_combined_2023} located a distorted structure with $a$=$b$=6.99 {\AA} by PBE-D3.\cite{grimme_consistent_2010} Calculations favoring planar g-C$_3$N$_4$\cite{khamdang_computational_2022, *zhu_tuning_2021, *yang_dft_2021, *liang_7.13_2021, *wang_7.13_2020, *tong_first-principle_2018, *sun_first_2018, *zuluaga_plane_2015} predicted almost the same lattice constants of 7.13 {\AA}. Energy stability was studied and the more stable corrugated than planar g-C$_3$N$_4$ from theoretical studies\cite{sehnert_saddlepoint_2007, liu_buckled_2020, azofra_dft_2016, wang_determination_2017} was attributed to the reduced steric repulsion from the pseudo-Jahn-Teller effect.\cite{ivanov_PJT_2015, ren_whether_2019, gracia_corrugated_2009, ahmed_energy-level_2022} Different catalytic (CO$_2$ conversion\cite{azofra_dft_2016}, N$_2$ reduction\cite{ren_whether_2019}) and optical\cite{ahmed_energy-level_2022} properties of both structures were compared.   

Additionally, a third structure (Fig. \ref{fig:periodicStructure}(c))\cite{zhang_accurate_2019} between  I and II was obtained in a hybrid way: from constraint geometrical optimizations with fixed cell length of $a$=$b$=6.80 {\AA}, taken from an XRD experiment.\cite{wang_6.81_2009} We refer to this \emph{micro-corrugated} structure as configuration III (Fig. \ref{fig:periodicStructure}(c)), where the central nitrogen in the heptazine unit (N3; Fig. \ref{fig:periodicStructure}(d)) is elevated above the molecular plane.

In this work, we present a comprehensive investigation of C1s and N1s XAS spectra at the three representative structures. Different spectral simulations are employed to validate and assess our prediction's accuracy. DFT and time-dependent DFT (TDDFT) are most commonly used for K-edge XAS simulations in materials.\cite{besley_modeling_2021, *besley_density_2020} With DFT, the full (FCH),\cite{TP1998} half (HCH),\cite{TP1998} or equivalent (ECH, or Z+1)\cite{Best.ECH, *JollyECH, *Davis.ECH,  *adams_equivalent_1993} core-hole approximation is often used. Multi-electron transitions are approximated as single-electron excitations from a core to virtual orbitals. In TDDFT,  core transitions are treated by using linear-response or real-time approaches.\cite{stener2003time, *fronzoni2005spin, *lopata2012linear} The Tamm-Dancoff approximation (TDA)\cite{hirata_tddft/tda_1999} is sometimes evoked for computational efficiency and to mitigate possible spin contamination issues.\cite{besley_tda_2016} Thus, three first-principles methods FCH, ECH, and TDDFT/TDA are selected. 

After the simulation, spectral interpretation is required to analyze the underlying electronic transitions for XAS peaks, helping uncover valuable physical/chemical insights.  While FCH-DFT can predict accurate energies, its MOs are not suitable for analyzing electronic transitions. In ECH-DFT calculations, the canonical molecular orbitals (ECH-CMOs) are often employed to aid the transition analysis.\cite{gao_chirality_2009, zhang_accurate_2019, ma_local_2019, ge_qmmm_2022, hua_x-ray_2010}  TDDFT calculations, with or without TDA, often rely on natural transition orbitals (NTOs)\cite{martin2003NTO} for interpreting the valence\cite{martin2003NTO, hua_multiple_2013, biggs_two-dimensional_2012} or core\cite{scutelnic_x-ray_2021, hua_multiple_2013, donahue_sulfur_2014} transitions. NTOs at different structures can help visualize the electronic transitions accompanied by structural dynamics.\cite{hua_monitoring_2016, hua_transient_2019, list_probing_2020} Hence, in this work, spectral analysis is conducted for ECH-DFT and TDDFT/TDA methods, allowing for a comparison.

The goal of this study is to offer possible explanations for the discrepancies in XAS of g-C$_3$N$_4$ from different experiments and to provide a theoretical reference for future studies involving more complex composite/derivative structures or transient structures. 


\begin{figure}
\centering
\includegraphics[width=8.0cm]{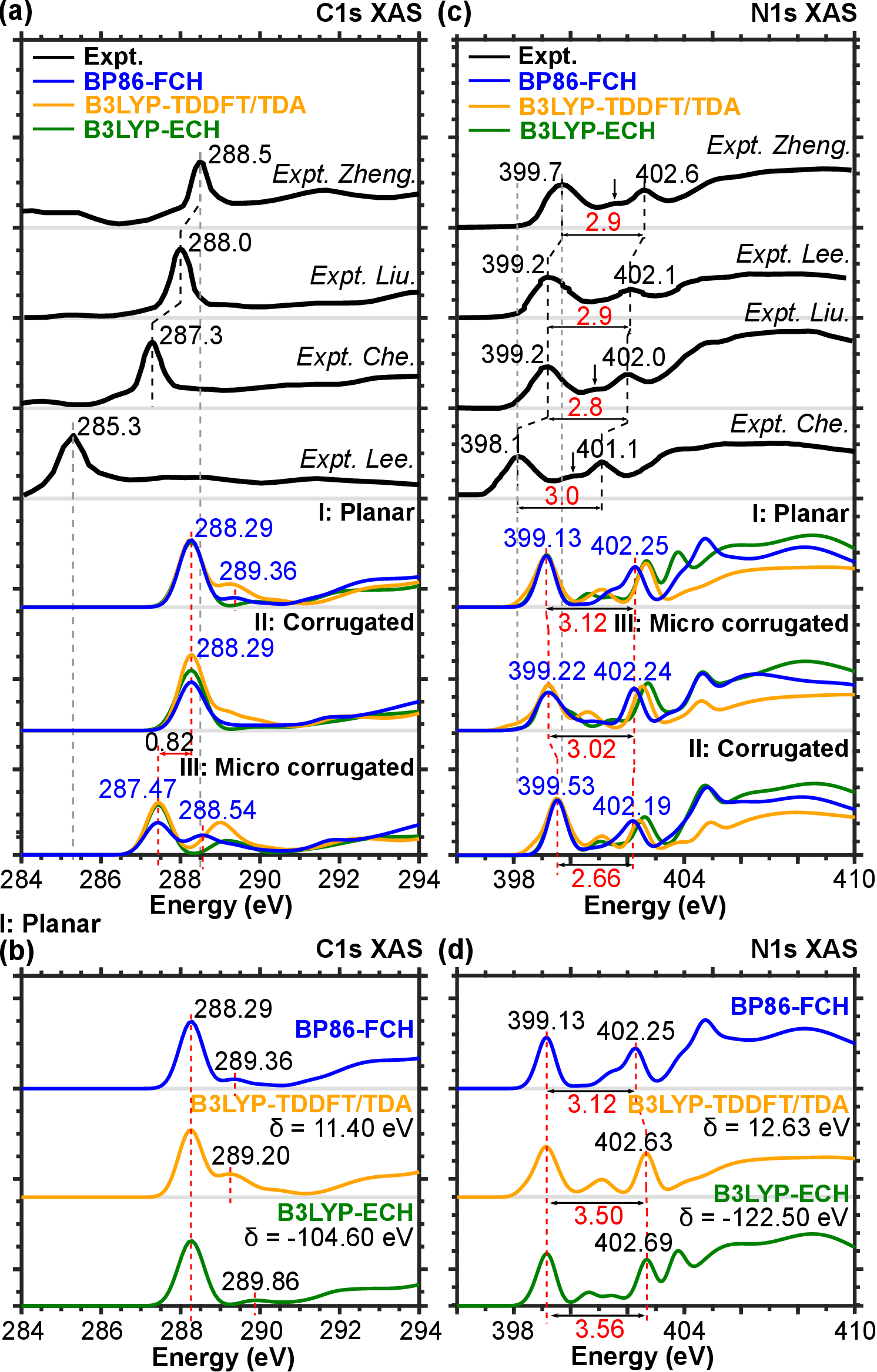}
\caption{(a) \textit{Bottom}: Simulated C1s XAS spectra of g-C$_3$N$_4$ configurations I-III by three methods. \textit{Top}: Experimental spectra from Zheng et al.,\cite{zheng2014hydrogen} Liu et al.,\cite{Liu_exp_2015} Che et al.,\cite{Che_exp_JACS2017} and Lee et al.\cite{Lee_exp2014} (b) A separated view of theoretical C1s spectra for I computed by three methods. (c-d) N1s edge results.}
\label{fig:XASthreeMethod} 
\end{figure}


\begin{figure}
\centering
\includegraphics[width=8.2cm]{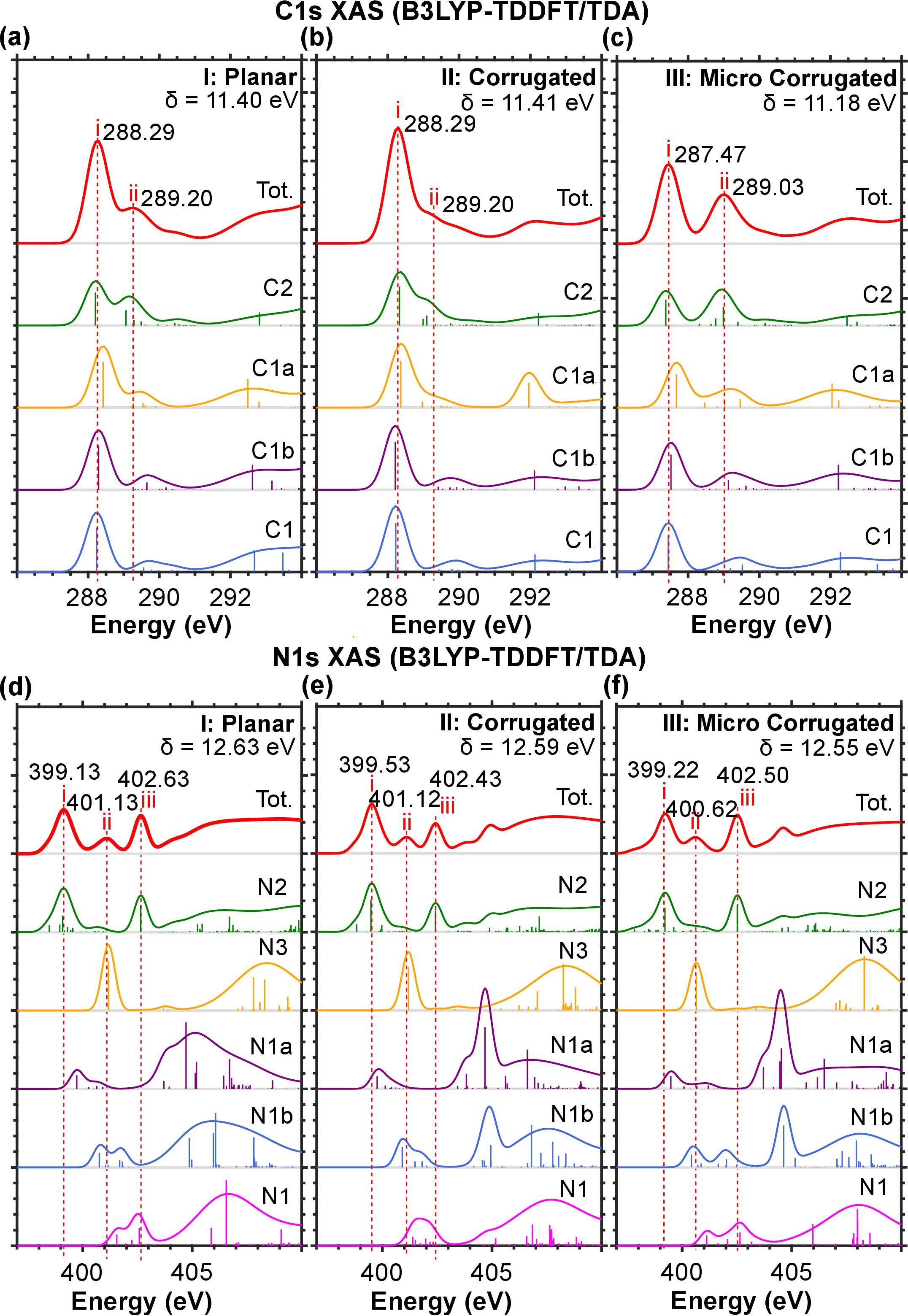}
\caption{Atom-specific contributions. Total (a-c) C1s and (d-f) N1s XAS of g-C$_3$N$_4$ configurations I-III simulated by B3LYP-TDDFT/TDA together with (unweighted) atom-specific contributions: (a,d) I, (b,e) II, (c,f) III. $\delta$ denotes the \textit{ad hoc} shift used.}
\label{fig:td} 
\end{figure}


\begin{figure*}
\centering
\includegraphics[width=14cm]{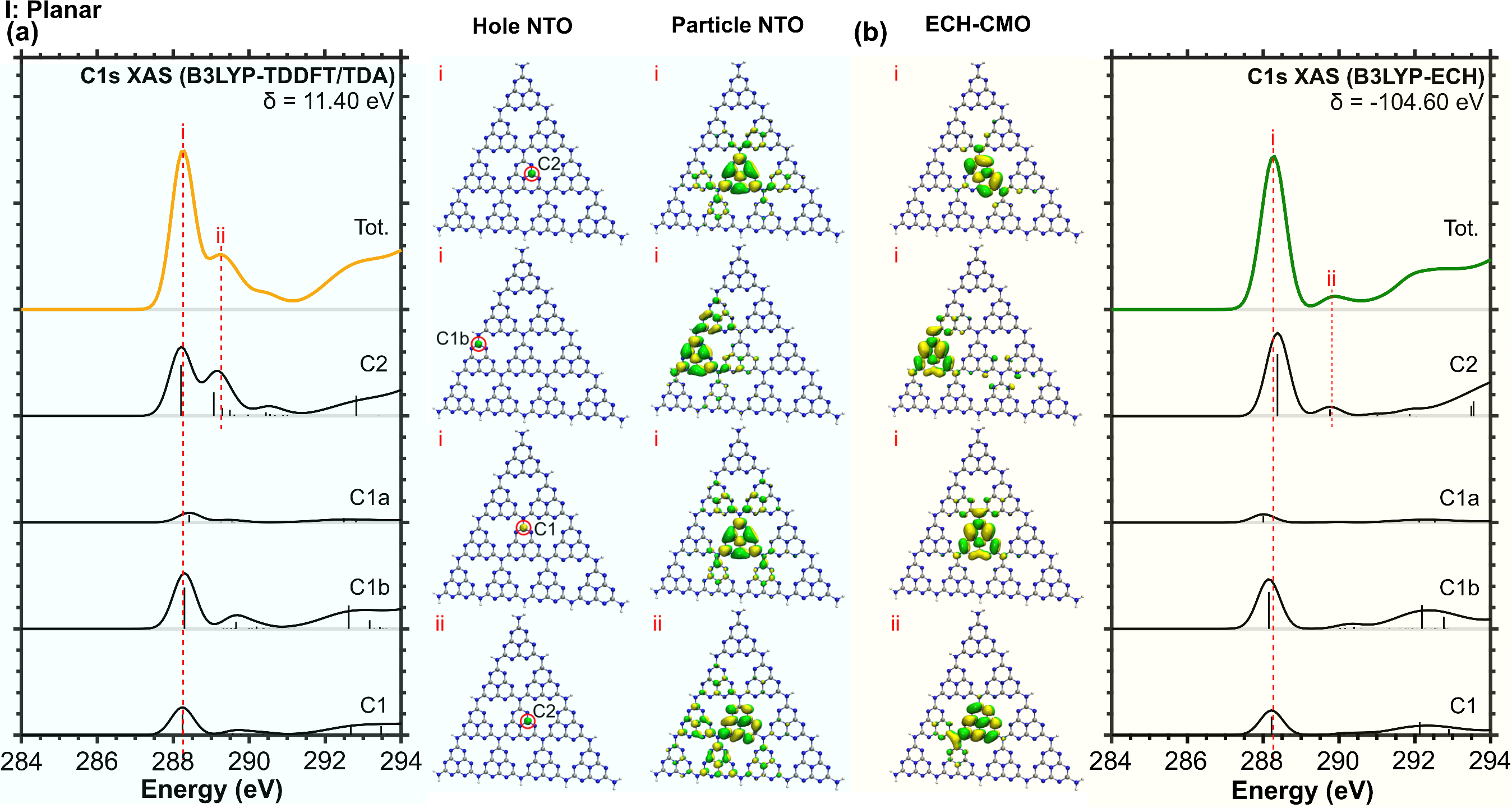}
\caption{Comparison of orbital analysis methods.
 Atom-specific C1s XAS spectra for g-C$_3$N$_4$ configuration I simulated by (a) B3LYP-TDDFT/TDA and (b) B3LYP-ECH methods.
 Major peaks are labeled, with corresponding  NTOs or ECH-CMOs shown beside the spectra. Circles represent the core centers.}
\label{fig:NTO:FSMO} 
\end{figure*}


Three different g-C$_3$N$_4$ structures, including planar (I), corrugated (II), and micro-corrugated (III) conformations, were selected for this study (Fig. \ref{fig:periodicStructure}(a-c)). In this work, structure I was relaxed by DFT with the PBE functional\cite{PhysRevLett.77.3865} using the VASP software.\cite{PhysRevB558} Structures of II and III, with bulking heights of 1.39 and 0.56 {\AA}, respectively, were sourced from our prior studies (II\cite{tian_NatCommun_2018} and III\cite{zhang_accurate_2019}). Both structures were optimized with the VASP software. Especially, relaxation of structure III had used the PBE functional\cite{PhysRevLett.77.3865}  and enforced fixed experimental\cite{wang_6.81_2009} cell length of 6.80 {\AA}.  While for Structure II, the optPBE-vdW functional\cite{dion2004van, klimevs2009chemical, klimevs2011van} had been employed with better consideration of the van der Waals (vdW) effect. A 160-atom cluster (Fig. \ref{fig:periodicStructure}(d)) was used to represent the periodic 2D material for XAS simulations. XAS spectra were simulated at the C1s and N1s edges using constructed cluster models for the three g-C$_3$N$_4$ structures. Each structure was analyzed with three different methods for comparison: FCH-DFT, ECH-DFT, and TDDFT/TDA. FCH and ECH calculations were performed using our PSSXS code\cite{PSSXS}, interfaced respectively to Q-Chem\cite{epifanovsky_Qchem5_2021} and Gaussian09\cite{g09} packages for electronic structure. TDDFT/TDA simulation used Q-Chem.\cite{epifanovsky_Qchem5_2021} More details are provided in supplementary note S1.



Our calculations show that the ground-state energies for structures I-III are 0.53, 0.00, and 1.71 eV (Fig. \ref{fig:periodicStructure}(a-c)). The corrugated structure II exhibits to be the most stable, consistent with previous theoretical studies.\cite{sehnert_saddlepoint_2007, liu_buckled_2020, azofra_dft_2016, wang_determination_2017} The micro-corrugated structure III has the highest energy owing to the imposed constraint of experimental\cite{wang_6.81_2009} cell lengths. Their corresponding lattice constants are $a$=$b$=7.13, 6.94, 6.80 {\AA}, respectively.  N1–C1 lengths display the largest differences of ca. 0.1 {\AA}  (1.48/1.42/1.38 {\AA} in  I/II/III), in alignment with the order of cell lengths. Differences in the remaining lengths are much smaller ($<$0.04\AA): C1–N2, N2–C2, and C2–N3 lengths are respectively 1.33/1.35/1.31, 1.33/1.34/1.30, and 1.39/1.40/1.39 {\AA} for I/II/III. 

Figure \ref{fig:XASthreeMethod}(a) compares the C1s XAS spectra from different experiments,\cite{zheng2014hydrogen, Liu_exp_2015, Che_exp_JACS2017, Lee_exp2014} revealing evident discrepancies (ca. 3 eV) in photon energies. Especially, the main peak position from the experiment by Lee et al.\cite{Lee_exp2014} (285.3 eV) stands apart from the other three\cite{zheng2014hydrogen, Liu_exp_2015, Che_exp_JACS2017} (287.3-288.5 eV). Interestingly, our simulated spectra at I-II (both at 288.3 eV) agree best with the experiments by Zheng et al.\cite{zheng2014hydrogen} (288.5 eV) in terms of the main peak energy; while our simulated spectrum at III (287.5 eV) best matches with the experiment by Che et al.\cite{Che_exp_JACS2017} (287.3 eV). Our calculations clarify a direct correlation between structural variations and the observed spectral discrepancies across these experiments. Additionally, among the theoretical spectra, the main peak of III manifests a redshift of 0.82 eV in comparison to I and II, possibly due to the constraint of a frozen lattice constant.

The spectral profiles of the three structures exhibit noticeable differences, with structure III showing a more pronounced second $\pi^*$ peak compared to  I and II. A separated view of the spectra at structure I computed by the three methods is shown in Fig. \ref{fig:XASthreeMethod}(b). The two DFT methods yield similar total profiles, while TDDFT/TDA predicts a significantly more intense $\pi^*$ peak.


Figure \ref{fig:XASthreeMethod}(c) recaptures recent experimental\cite{zheng2014hydrogen, Lee_exp2014, Liu_exp_2015, Che_exp_JACS2017} N1s XAS spectra of g-C$_3$N$_4$ compared with our simulated results at configurations I-III. Regarding the main peak energy, only one experiment by Che et al.\cite{Che_exp_JACS2017} (398.1 eV) deviates from the other three (399.2-399.7 eV).\cite{zheng2014hydrogen, Lee_exp2014, Liu_exp_2015} Our calculations at  I (399.1 eV) and II (399.2 eV) coincide with the experiments by Lee et al.\cite{Lee_exp2014} and Liu et al.\cite{Liu_exp_2015} (both at 399.2 eV), while our result at III (393.5 eV) generates main peak energy between values reported by Lee et al.\cite{Lee_exp2014}/Liu et al.\cite{Liu_exp_2015}  (399.2 eV) and Zheng et al. (399.7 eV).\cite{zheng2014hydrogen} Similar to the C1s edge, our calculations validate the accuracy by agreement with the experiments and demonstrate the effects of structural variations on the main peak shift. II exhibits a blue shift of 0.3-0.4 eV relative to the other two. The separation between the two main $\pi^*$ peaks is essentially consistent in  I (3.12 eV) and III (3.02 eV), but much smaller in II  (2.66 eV). It is noted that the range of 2.66-3.12 eV is consistent with the experimental gap of 2.8-3.0 eV.\cite{zheng2014hydrogen, Liu_exp_2015, Che_exp_JACS2017, Lee_exp2014} In summary, the significant corrugation in II leads to a distinct blue shift in the energy, alongside a narrowing of the gap between the two $\pi^*$ peaks in the N1s XAS. 

When analyses at both the C1s and N1s edges [Fig. \ref{fig:XASthreeMethod} (a) and (c)] are ready, the ideal scenario is that both edges yield the same results. This alignment is evident for configurations I and II. However, for III, while a good agreement with the experiment by Che et al.\cite{Che_exp_JACS2017} at the C1s edge was observed, there was a deviation of ca. 1.1 eV in the absolute transition energy at the N1s edge. To ensure consistent assignments, it is essential for experiments to be conducted under the same conditions with consistent calibrations and for simulations to maintain equal accuracy at both edges. The computational precision achieved with the DFT-based $\Delta$KS scheme is generally considered to have sub-electronvolt accuracy for absolute core energies.\cite{takahashi_functional_2004, pueyo_bellafont_validation_2015, pueyo_bellafont_performance_2016, bagus_consequences_2016, kahk_accurate_2019, du_theoretical_2022} The specific numerical values, though, can vary depending on the systems and edges, exhibiting both underestimation and overestimation. With current DFT-based computational techniques, it is difficult to guarantee identical deviations at both the C1s and N1s edges. It is commonly believed that relative energies among different structures at the same edge have higher accuracy, as often assumed in general chemical reaction studies using quantum chemistry. Despite some sub-eV uncertainty in the \textit{absolute} energies of the main peaks in configurations I-III (399.13, 399.22, and 399.53 eV), the range of \textit{relative} energy (spanning 0.4 eV) is deemed more reliable (Fig. \ref{fig:XASthreeMethod}(c)).  However, the experimental N1s absorption energy reported by Che et al.\cite{Che_exp_JACS2017} appears to deviate significantly (1.1-1.6 eV) from the other three experiments. Nevertheless, the N1s spectral profiles of configuration III and the experiment by Che et al.\cite{Che_exp_JACS2017} exhibit good agreement, consistent with the correspondence observed at the C1s edge. One can find that the $\pi^*$ separations in III and the experiment by Che et al.\cite{Che_exp_JACS2017} are nearly identical (3.0 eV).

Figure \ref {fig:XASthreeMethod}(d) compares the three methods applied to the same structure, showing essentially similar profiles but with evident differences. The separation between the two $\pi^*$ peaks is 3.12 (FCH), 3.50 (TDDFT/TDA), and 3.56 (ECH) eV, respectively. 


Figure \ref{fig:td} provides further analysis for different atomic sites, taking the TDDFT/TDA results as an example. At each structure, the main peak positions of different carbon sites (C1, C1a, C1b, and C2) in the C1s XAS spectra almost coincide with each other (Fig. \ref{fig:td}(a-c)). It is now more clear that the differences across I-III mainly come from the spectrum of C2. All three structures give two main $\pi^*$ peaks by C2, differing by their relative energies and intensities. Especially,  III exhibits two more separated (1.6 eV), $\pi^*$ peaks with almost equal intensities in its C2 spectrum, while for the other two configurations, the separation between the two peaks is much smaller (0.9 eV)  and the lower-energy peak is much stronger than the higher-energy one. Thus, C2 takes the major responsibility for the discrepancies in the total spectra.  Structurally, C2 is bonded to N3  (Fig. \ref{fig:periodicStructure}(d)). In III, N3 (Fig. \ref{fig:periodicStructure}(c)) has a more elevated position (as referred to the molecular plane) than other atoms. Hence,  it exhibits evident changes in the spectral profiles. Note that the micro-corrugation at N3 not only influences C2 but can also cause a global redshift of 0.8 eV to the main peak positions of all carbons. Although structure II generally has a larger corrugation, its corrugation seems to be delocalized over all atoms. 

C2 weights 50\% to the total spectra, and C1, C1a, and C1b make the remaining 50\%. Structurally, C1a and C1b can be considered as C1 at the edge bonded with terminal -NH$_2$ and -NH groups, respectively (Fig. \ref{fig:periodicStructure}(d-e)). Analyzing the atom-specific spectra allows us to examine the influence of terminal hydrogens.  They show substantial similarity, implying that the terminal -NH$_x$ bond exerts a minimal effect on the C1 spectra. A small blue shift ($<$0.3 eV) can be identified in the sequence from C1 (C-N$_3$), C1b (C-NH), to C1a (C-NH$_2$).


Figure \ref{fig:td}(d-f) shows the atomic-specific contributions to the total N1s XAS. Structural variations exert less influence compared with the C1s edge. Among the nitrogens, N2 has the largest weight (75\%) and exerts the dominant influence on the total spectra.   It determines the energy of the main peak and the separation between the two main $\pi^*$ peaks.  For structures I--III, the N2 spectra exhibit main peak positions at 399.1, 399.5, and 399.2 eV, respectively.  N3 weights 12.5\%. It contributes similar profiles but evident differences in photon energies among configurations, leading to a distinct weak signature (400.6 eV, III; 401.1 eV, I--II) between the two main $\pi^*$ peaks in the total N1s spectra,  characterized by an energy difference of ca. 0.5 eV.  The remaining nitrogens (N1, N1a, N1b) collectively give a total weight of 12.5\% and make minor contributions to the total spectra.


To compare different orbital analysis methods from TDDFT/TDA and ECH calculations, theoretical C1s XAS spectra at configuration I are interpreted in Fig. \ref{fig:NTO:FSMO}. Corresponding NTOs and ECH-CMOs are also shown. The particle NTOs can be compared with these ECH-CMOs.  Generally,  orbitals generated from both methods lead to consistent interpretations, localizing at similar regions and having close shapes. In the N1s edge (Fig. S2), the analyses conducted using both methods reveal generally consistent graphics (supplementary note S2). When the same method (ECH) is used, different structures are reflected in changes in ECH-CMOs (Figs. S3-S4), showcasing the potential of ECH orbitals for future transient studies of materials. This is similar to how the evolution of NTOs was  used previously to identify the ultrafast dynamics of gas-phase molecules.\cite{hua_transient_2019, hua_monitoring_2016} 

In summary, we computed  XAS spectra of three g-C$_3$N$_4$ configurations using three first-principles methods to study the effects of structural variations and the method performances. Within each method, the spectra exhibited noticeable differences among the three structures, which can be related to different experiments.  Our calculations explained that the observed discrepancies in experiments may arise from varying structures obtained during synthesis or under different ambient conditions. Despite visible differences, simulated spectra from the three methods consistently predict the fundamental spectral features. Sticking with a single fixed method can study structural dynamics and time-resolved X-ray spectra. Additionally, we compared two orbital analysis methods, NTOs (by TDDFT/TDA) and ECH-MOs (by ECH), confirming their consistency. Both methods can assist in peak characterization and spectral interpretation. The sensitivity of the structure-orbital relationship highlights the potential of these interpretation methods for future time-resolved XAS studies. ~\\

See the \textcolor{blue}{supplementary material} for computational details, additional orbital analyses, and supplementary Figs. S1-S4.

Financial support from the National Natural Science Foundation of China (Grant No. 12274229) is greatly acknowledged.

All computational details supporting the findings of this study are available within the main text and the supplementary material. The data are also available from the corresponding authors upon reasonable request.

%

\end{document}